\documentclass[twocolumn]{aa}
\usepackage{graphicx}
\usepackage{txfonts}
\begin{document}
   \title{A comparison of density structures of a star forming and a non-star-forming globule \thanks{Based on observations collected at the European Southern Observatory, Paranal, Chile (ESO Programme  68.C-0318)}}
   \subtitle{\object{DCld303.8-14.2} and \object{Thumbprint Nebula}}

   \author{J. Kainulainen\inst{1 \and 2} \and K. Lehtinen\inst{2} \and P. V\"ais\"anen\inst{3} \and L. Bronfman \inst{4} \and J. Knude \inst{5}
          }

   \offprints{J. Kainulainen \email{jkainula@astro.helsinki.fi}}

   \institute{European Southern Observatory, Karl-Schwarzschild-Str. 2, D-85748 Garching bei M\"unchen\\
              \email{jkainula@astro.helsinki.fi}
   \and Observatory, P.O. Box 14, SF-00014 University of Helsinki \\
              \email{lehtinen@astro.helsinki.fi}
   \and South African Astronomical Observatory, P.O. Box 9, Observatory, 7935, South Africa \\
              \email{petri@saao.ac.za}
   \and Universidad de Chile, Departamento de Astronom\'ia, Casilla 36-D, Santiago, Chile  \\
              \email{leo@das.uchile.cl}
   \and Niels Bohr Institute, Juliane Maries Vej 32, DK-2100, Copenhagen \O, Denmark \\
              \email{indus@astro.ku.dk}}

   \date{Received <date> ; accepted <date>}

\abstract{We present a study of radial density structure of the star forming globule, DCld303.8-14.2 (DC303), and a non-star forming globule, Thumbprint Nebula (TPN), using near-infrared data taken with the ISAAC instrument on the Very Large Telescope. We derive the extinction through the globules using the color excess technique and examine the radial density distribution using Bonnor-Ebert and power-law models. The two globules have significantly different density structures. The extinction profile of DC303 is best fitted with a single power-law with an exponent $p = 2.29 \pm 0.08$. An unstable Bonnor-Ebert model with a dimensionless parameter $\xi_\mathrm{max}= 23 \pm 3$ provides equally good fit to data. The extinction profile of TPN flattens at small radii, making the profile significantly different from the profile of DC303. We are unable to fit the Bonnor-Ebert model for TPN in a robust manner, but derive the lower limit $\xi_\mathrm{max}\gtrsim 8$ for the dimensionless outer edge. The density profile derived for TPN is typical compared to recently observed pre-protostellar globules, with high  $\xi_\mathrm{max}$ value which could be interpreted as the presence of significant additional support or very slow contraction.

   \keywords{Stars: formation - ISM: clouds - ISM: individual objects: DCld303.8-14.2 - ISM individual objects: Thumbprint Nebula
               }

   }

\titlerunning{A comparison of a star forming and a non-star-forming globule}

\authorrunning{Kainulainen et al.}

   \maketitle
%

\section{Introduction}

Small-sized clouds of cold interstellar matter, the globules, are often referred to as ideal places to study processes and phenomena connected to low-mass star formation. The ideality arises from the apparent simplicity of these objects: they are nearly spherical, isolated, slowly evolving clouds and only a small number of stars, if any, can be formed in their interiors. The sizes of typical globules are in the order of tenths of parsecs, their masses range from $\sim$1 M$_\odot$ \dots 20 M$_\odot$, and the temperature in them have been observed to be around 10 K. Roughly one third of the globules contain point sources, indicating the presence of a newly born star (Lee \& Myers \cite{lee99}).

Once a globule has been formed in a molecular cloud it may evolve towards a point where the gravitational force inevitably takes over and collapse ensues. In theories of star formation the properties of the entire collapse process depend on the characteristics of the cloud at the onset of the collapse, i.e. at the \emph{initial stage} of the collapse. Factors such as mass distribution, temperature, velocity and magnetic fields at initial stage predict the future evolution of the collapsing core. Some of the factors, like temperature and large scale velocity fields, can be well traced with observations of molecular lines. Further, the significance of some factors, e.g. magnetic fields, are not so well understood due to observational difficulties. 

In particular, theoretical star formation scenarios predict the density distribution for the core at the initial stage of the collapse. The form of the density distribution is linked to the rate at which a protostar accretes mass from the surrounding envelope and further to the timescale of the collapse. It is thus an essential factor considering the process of star formation. In the classical 'inside-out' collapse scenario of Shu (Shu \cite{shu77}, S77 hereafter) the initial stage of core collapse equals to the singular isothermal sphere with a steep density distribution $\rho (r) \sim r^{-2}$. After this configuration is established the collapse begins in the center of the core expanding ``inside-out'' as a shock-front. The mass accretion rate in S77 is constant throughout the collapse for all radii. Some other models, such as those of Foster \& Chevalier (\cite{foster93}) who take a marginally stable Bonnor-Ebert sphere as an initial condition, or Ciolek \& Mouschovias ({\cite{ciolek94}) and Basu \& Mouschovias (\cite{basu94}) who take ambipolar diffusion and magnetic fields into account, assume \emph{flat} inner density profiles for the initial stage of the collapse. In those models the accretion rate is a function of both radius and time.

The observations of pre-stellar cores having both $\rho (r) \sim r^{-2}$ and flat inner profiles indeed have been made using sub-millimeter dust continuum observations (e.g. Ward-Thompson et al. \cite{ward-thompson94}, \cite{ward-thompson99}; Chandler \& Richer \cite{chandler00}; Shirley \cite{shirley00}; Johnstone et al. \cite{johnstone00}, \cite{johnstone01}, \cite{johnstone05}; Evans et al. \cite{evans01}; Kirk et al. \cite{kirk05}, \cite{kirk06}). For some individual globules the mass distribution have been obtained via molecular line observations and chemical modeling (e.g. Zhou et al. \cite{zhou93}; van der Tak et al. \cite{tak05}). However, derivation of the density distribution using these methods, i.e. interpreting the observations in terms of density, suffers from the unknown properties of the selected model: excitation conditions, possibly large optical depths, time dependent chemistry or unknown temperature and emissivity of the dust grains. 

The extinction of background starlight caused by dust particles provides a more direct method than those above for determination of the density distribution. The colors of the stars shining through the cloud are measured at two or more bands and the colors are compared to the colors of stars in a nearby reference field. The near-infrared is particularly suitable for the observations as the extinction is about one tenth of the extinction in visual wavelengths and cores therefore more transparent. This method has been successfully used in the context of dark globules, leading for example to the suggestion that pre-protostellar cores could be well described by pressure confined Bonnor-Ebert models close to the hydrostatic equilibrium (Alves et al. \cite{alves01}). Moreover, protostellar cores which show strong evidence of star formation activity have been observed to have density distributions corresponding to highly unstable Bonnor-Ebert spheres or steep power-laws with exponent $p \gtrsim 2$ (Harvey et al. \cite{harvey01}; Harvey et al. \cite{harvey03}).

The observations of density distributions from both pre-protostellar and protostellar cores can be used to seek the characteristics of different star formation scenarios. In this paper we exploit the near-infrared color excess technique to study the radial density distributions of two cloud cores: we determine the mass distribution of a core that has very recently formed a star, and a core which is regarded to be quiescent, close to hydrostatic equilibrium and thus possibly close to the initial stage of star formation. For this purpose we have selected two globules, namely DCld303.8-14.2 (or Sandquist 160) and Thumbprint Nebula, and conducted deep near-infrared $JHK$ surveys towards them. 

The two globules are introduced in \S\ref{sec_objects}, and in \S\ref{sec_observations} we describe the observations, data reduction and the photometry. The radial density distributions of the globules are presented and discussed in \S\ref{sec_results} and \S\ref{sec_discussion}. The conclusions of the work are presented in \S\ref{sec_conclusions}.


\section{DCld303.8-14.2 and Thumbprint nebula}
\label{sec_objects}

The globule DCld303.8-14.2 (DC303) and Thumbprint Nebula (TPN) are both members of the nearby, quite extensively studied molecular cloud complex of Chamaeleon (Cha). They are located close to the clouds Cha II and Cha III, from which Cha II is an active star forming cloud with tens of Young Stellar Object (YSO) candidates within its boundaries and Cha III is a quiescent cloud. The two globules, DC303 and TPN, were initially selected for this study due to their similar morphological appearance and accessible location in a well-studied cloud complex.

Fig. \ref{fig_region-images} shows the IRAS 100 $\mu$m map of Cha II/III region and the Digitized Sky Survey blue plate images of both globules. The globules are separated by about 2\degr. Both cores are clearly visible in the 100 $\mu$m map and DC303 is actually one of the brightest 100 $\mu$m sources in Cha II after the central star forming region. The globules show very similar morphology in the DSS blue plate images (Fig. \ref{fig_region-images}): the surface brightness of scattered light forms a bright rim around the globule with maximum intensity on the northern side, which is the direction of the galactic plane. Inside the rim the surface brightness drops forming a dark center for the globules.


DC303 harbors an IRAS point source \object{IRAS 13036-7644}. The millimeter wavelength emission lines of $^{12}$CO($J$=1-0), CS($J$=2-1) and ($J$=5-4), HCO$^+$($J$=1-0), HCN($J$=1-0) by Lehtinen (\cite{lehtinen97}) and emission lines of H$_2$CO(2$_{12}$-1$_{11}$), C$_3$H$_2$(3$_{12}$-2$_{21}$) and CH$_3$OH(3$_0$-2$_0$ A$^+$) by Mardones et al. (\cite{mardones97}) have revealed a low-velocity outflow driven by the IRAS source. Also, some molecules show blue-skewed line profiles indicating the possibility of collapse motions. On the other hand, some lines such as HCO$^+$ show red-skewed profiles, and radiative transfer modeling would be required to further interpret the molecular line observations. The spectral energy distribution of IRAS 13036-7644 is typical of a very young protostar. Based on the characteristics of the SED and qualitative arguments about the molecular line profiles, the source is placed in the midway of Classes 0 and I in the traditional classification scheme of YSOs (Lehtinen et al. \cite{lehtinen03}, \cite{lehtinen05}; see e.g. Lada \& Wilking \cite{lada84}; Adams et al. \cite{adams87}; Wilking et al. \cite{wilking89} for the classification scheme).


The Thumbprint Nebula is an apparently quiescent globule with no detected infall/outflow motions or embedded point sources. It has been studied previously in near-infrared to constrain dust model parameters (Lehtinen et al. \cite{lehtinen96}) and in far-infrared to derive the dust temperature (Lehtinen et al. \cite{lehtinen95}; Lehtinen et al. \cite{lehtinen98}). The millimeter wavelength molecular lines of $^{12}$CO($J$=1-0) and ($J$=2-1), $^{13}$CO($J$=1-0) and ($J$=2-1), C$^{17}$O($J$=1-0), CS($J$=2-1) and HCN($J$=1-0) have been observed towards the globule, from which a very low kinetic temperature $T_{kin}\approx 6.6$ K was derived. The nebula also seems to be close to the thermodynamic equilibrium with $E_{pot}/E_{kin} \approx 0.9$ (Lehtinen et al. \cite{lehtinen95}). Lehtinen et al. also find that the energy balance of the globule is explained solely by the external radiation from the interstellar radiation field.

   \begin{figure*}
   \centering
   \includegraphics[width=0.65\columnwidth]{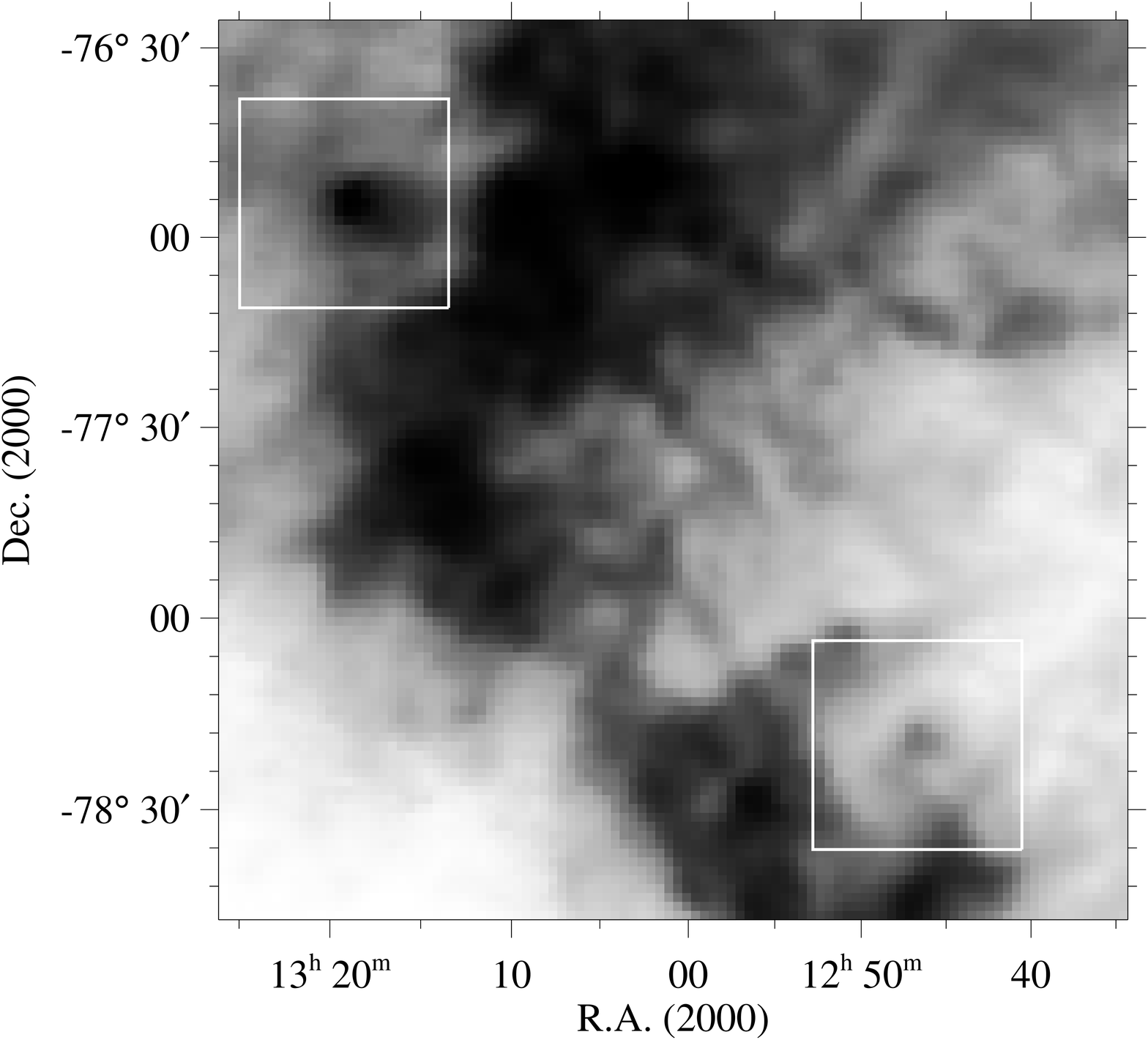} \includegraphics[width=0.65\columnwidth]{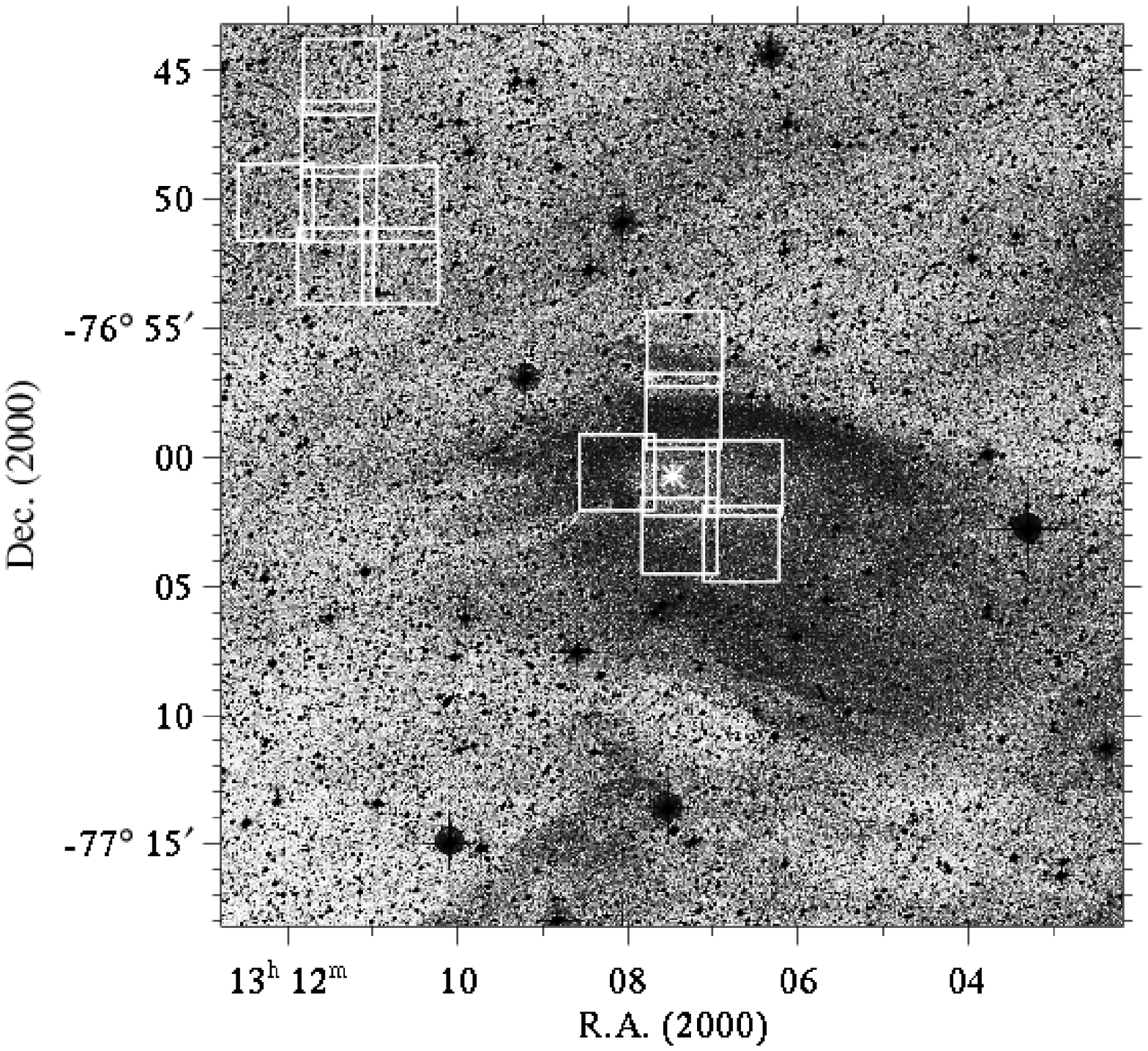} \includegraphics[width=0.65\columnwidth]{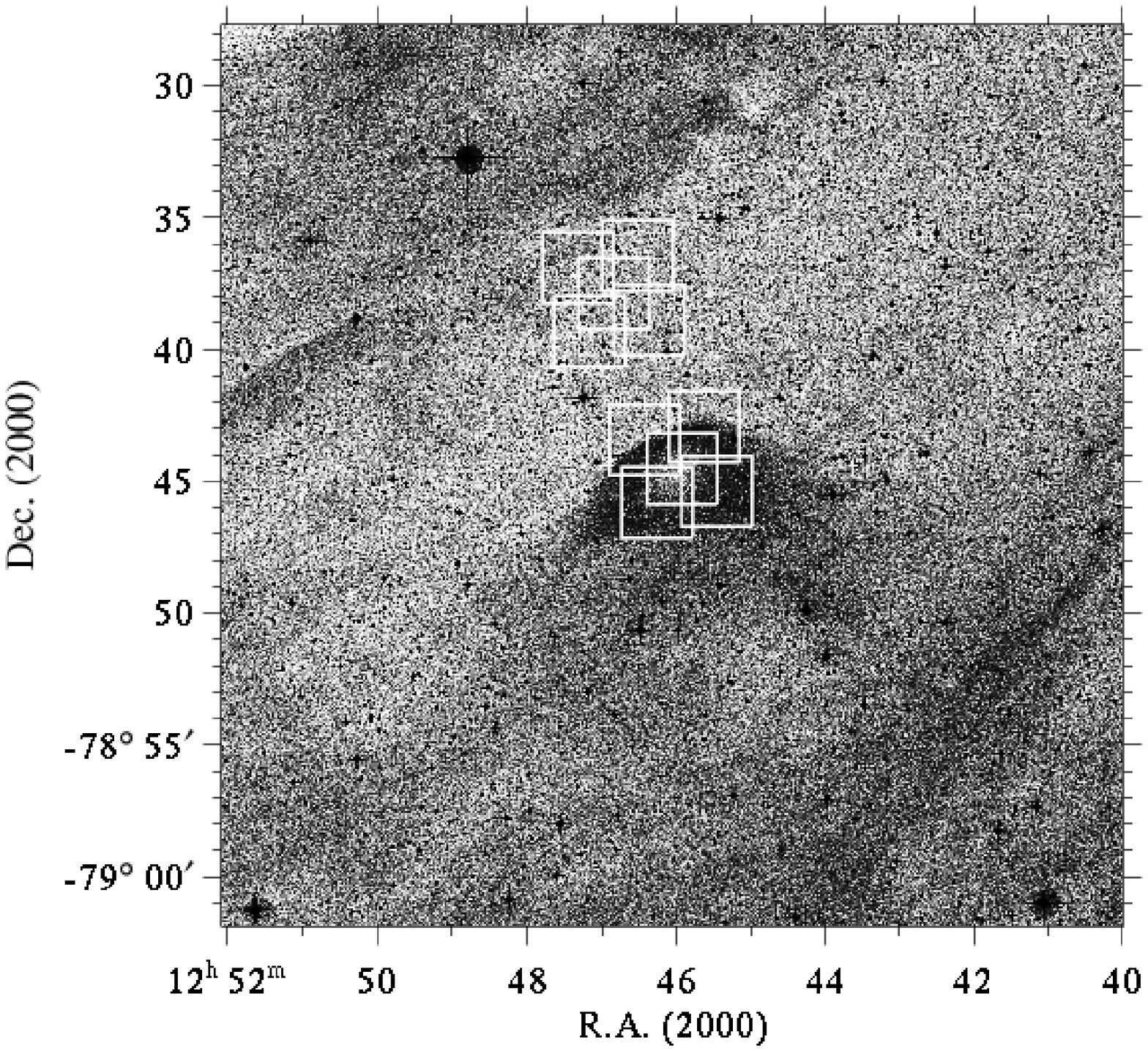}
      \caption{{\bf Left: }IRAS 100 $\mu$m emission map of the Chamaeleon II-III region. The white frames show the positions of the DSS images on the right. {\bf Center: }The Digital Sky Survey blue band image of the DC303 region. The frames observed with the VLT/ISAAC are marked with white rectangles. The position of the point source IRAS 13036-7644 is marked with an asterisk. {\bf Right: }The same for the Thumbprint Nebula.}
         \label{fig_region-images}
   \end{figure*}


\section{Observations and data reduction}
\label{sec_observations}

The observations were made in service mode with the ISAAC
(Infrared Spectrometer And Array Camera, Moorwood et al. \cite{moorwood98}) at VLT (Very Large
Telescope) during January-April 2002, in $J$, $H$ and $K_s$
bands.

In addition to the measurement of extinction, the objective of the observations was to detect the weak surface brightness of the scattered light. Because the extent of the scattered light
was expected to be about the same as the field of view of a
single image frame, the observations were made in an
OFF-ON-OFF$\dots$-ON-OFF sequence, where the OFF-position is
outside the cloud. Individual ON and OFF frames were taken as
30 sec or 1 min exposures. The OFF-positions were then used to derive the
sky background image. We observed several ON positions for both
clouds to fully map the clouds and to have some background sky
around the clouds. The positions of the ON and OFF fields are shown
in Fig. \ref{fig_region-images} for both globules.

The basic data reduction, such as sky substraction and flat-fielding, has
been performed with the xdimsum package within IRAF. To correct for the 
variability of the sky background, the 3-6 OFF-positions closest (in time) to each dithered ON-position were combined and substracted. These sky-substracted ON-frames were then shifted and added to create a total ON-position image of typically 20-30 min total exposure time. 

Then the SExtractor (Bertin \& Arnouts \cite{bertin96}) package was used 
to find point sources in the images.  The magnitudes of the stars have
been derived with the IDL version of the DAOPHOT aperture
photometry package (http://idlastro.gsfc.nasa.gov). The limitting magnitudes, defined as magnitudes whose photometric error is equal to 0.1 mag, are: $J=20.8$, $H=19.2$ and $K_\mathrm{S}=18.8$ mags for DC303 and  $J=20.7$, $H=19.0$ and $K_\mathrm{S}=18.7$ for TPN. Typical errors of $J$, $H$ and $K_\mathrm{S}$ band magnitudes are 0.05\,mag, 0.08\,mag and 0.08\,mag, respectively. 

The photometric calibration was derived using stars from the Persson faint NIR standard list (Persson et al. \cite{persson98}). Several photometric nights during the observations were used to fix the photometric scale, and the zero-points for the non-photometric nights were derived by utilizing overlapping fields from different nights. Overall photometry is uniform over the whole field within 5\%, and the $JHK_\mathrm{S}$ magnitudes were also checked to be consistent with 2MASS photometry within few percents. The photometry of all detected sources will be made available online at the CDS.


\section{Results}
\label{sec_results}

\subsection{NIR images of DC303}

The reduced center frames of DC303 observed with ISAAC at $K_\mathrm{S}$, $H$ and $J$ bands are shown in Fig. \ref{fig_images}. The images show two distinctive features: the bright circle of scattered light forming a circular halo around the center of the core and a number of non-pointlike sources close to the location of the embedded IRAS source (see also Lehtinen et al. \cite{lehtinenESO}).

The scattered light has been commonly observed towards dark clouds at optical wavelenghts, but applications to near-infrared wavelengths have been few. Dedicated observations of scattered light in near-infrared have been made to constrain dust models in dark clouds (Lehtinen \& Mattila \cite{lehtinen96}; Nakajima et al. \cite{nakajima03}), but the phenomenon has not been exploited in the context of tracing the mass distributions. Several studies in which NIR imaging of dark cores have been conducted have, however, reported the presence of diffuse structures (Lehtinen et al. \cite{lehtinenESO}; Kandori et al. \cite{kandori05}). Recently, Foster \& Goodman (\cite{foster06}) and Padoan et al. (\cite{padoan06}) studied the feasibility of using measurements of scattered light in near infrared to derive the column density distribution of a dark cloud and concluded that it can be used, when complemented with a proper radiative transfer modeling, as a new high-resolution tool. The measurements of scattered light seem to be particularly well suited to determine the density distribution of more diffuse clouds, in which the maximum extinction remains below $\sim 20$ mag (Juvela et al. \cite{juvela06}). Although the extinction in the center of DC303 is clearly higher, it offers an interesting ground to implement the treatment into the case of dark globules. Our aim is to model the scattered light to constrain the mass distribution and study the dust properties in DC303 in a forthcoming paper. The extended structures close to the IRAS source resemble the appearance of Herbig-Haro objects often seen associated with outflows from young stars. We have performed near-infrared long-slit spectroscopy of these objects to verify their nature.

   \begin{figure*}
   \centering
   \includegraphics[width=0.65\columnwidth]{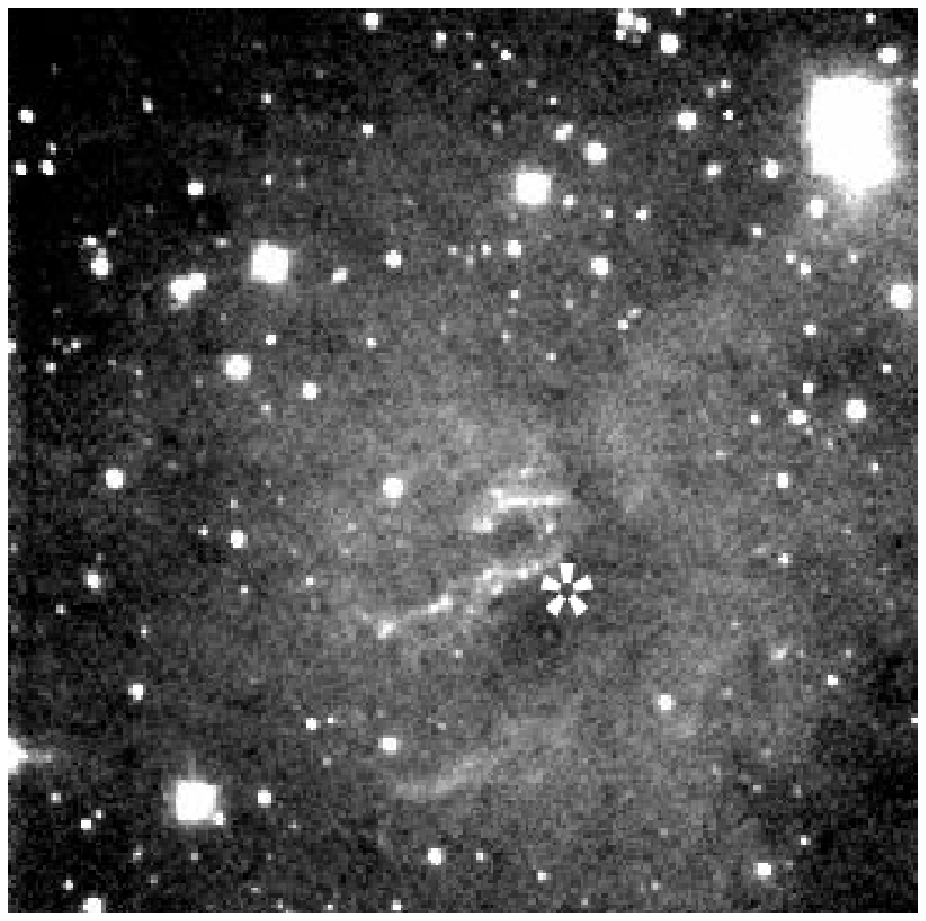} \includegraphics[width=0.65\columnwidth]{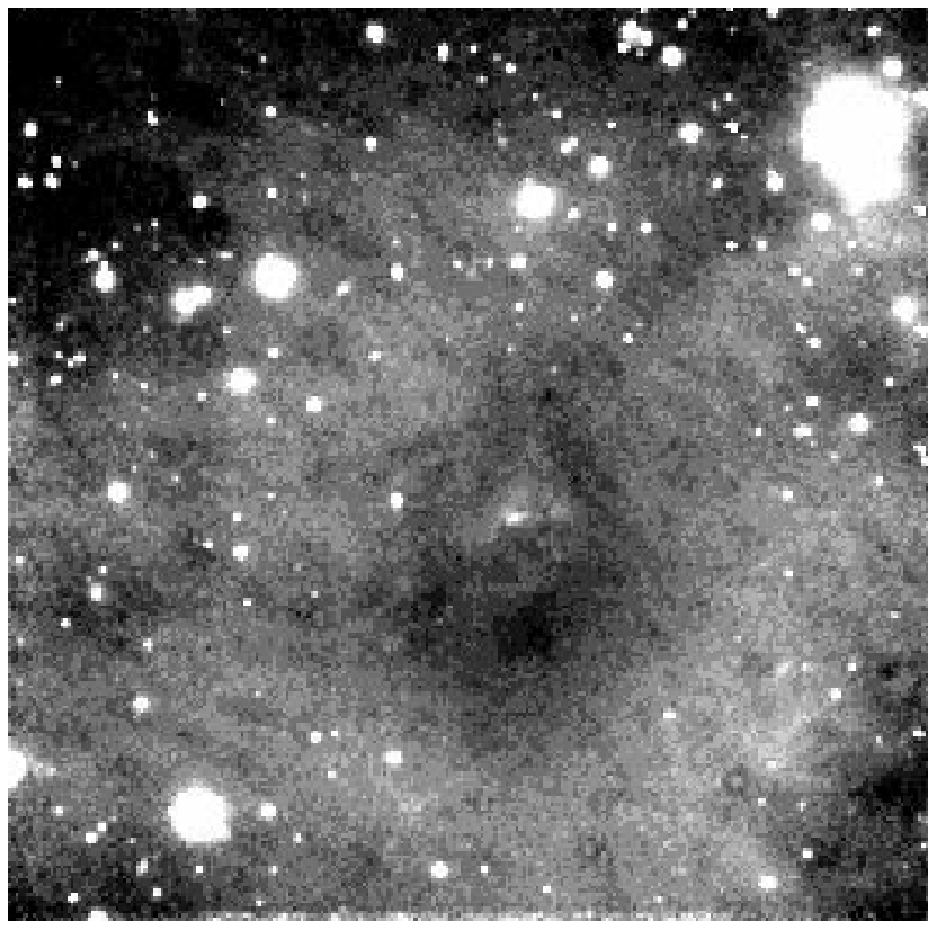} \includegraphics[width=0.65\columnwidth]{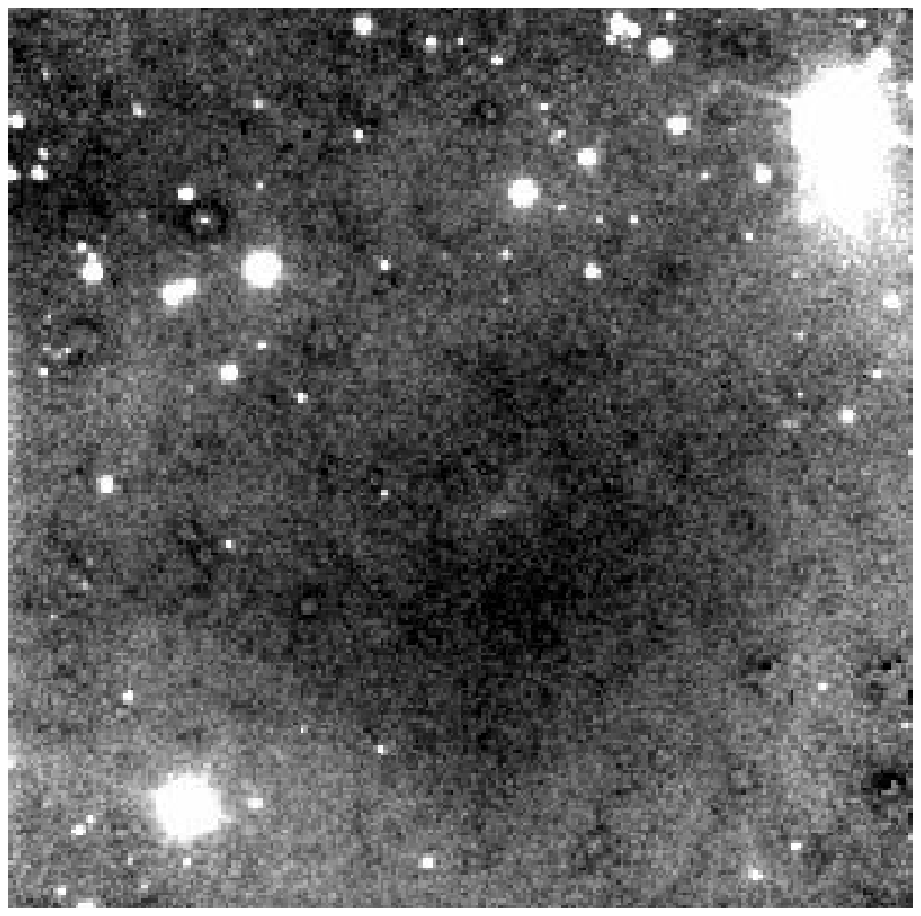} 
      \caption{The $K_\mathrm{S}$, $H$ and $J$ band images of DC303. The asterisk marks the position of the IRAS 13036-7644.}
         \label{fig_images}
   \end{figure*}


\subsection{NIR colors and extinction}

We use the $H$ and $K$ magnitudes of the observed stars to derive the near-infrared extinction in the globules. In particular, we use the color excess technique described by Lada et al. (\cite{lada94}, ``NICE''). The color excess along the line of sight towards a star is related to the observed magnitudes through the equation:
\begin{equation}
E(H-K)=(H-K)_\mathrm{obs} - (H-K)_\mathrm{int},
\label{eq_color-excess}
\end{equation}
where $(H-K)_\mathrm{int}$ is the intrinsic color of the star. In the NICE method, this equation is solved by assuming that the intrinsic colors of background stars are, on the average, similar to the colors of stars observed on a nearby field which is free from extinction. Furthermore, the color excesses can be converted to visual extinction by assuming a certain form of the reddening law, i.e. the ratio between color excess and visual extinction. Using e.g. the often quoted reddening law of Rieke \& Lebofsky (\cite{rieke85}) we obtain:
\begin{equation}
A_\mathrm{V}=15.87 \times E(H-K)
\label{eq_mathis}
\end{equation}
This conversion is not necessary for the derivation of radial density distributions in Sec. \ref{sec_rho-models}, but is provided to have a common ground for comparison with the results of other papers.

The average intrinsic colors calculated from the reference fields (Fig. \ref{fig_region-images}) are: $(H-K_\mathrm{S})_0=0.14 \pm 0.13$ for DC303 and  $(H-K_\mathrm{S})_0=0.13\pm0.09$ for TPN. Applying Eqs. \ref{eq_color-excess} and \ref{eq_mathis}, and using these intrinsic colors, an estimate is obtained for extinction towards each observed star in the globule fields. We assume that all the stars are background stars (i.e. there are no foreground stars). To produce a spatial distribution from the discrete data, the color excess values are then convolved using a gaussian function with 40$''$ and 30$''$ FWHM for DC303 and TPN, respectively. Each pixel in the final map is required to have at least one star within the FWHM radius from the pixel center to make the extinction calculation applicable. If there are no stars within that range, the extinction in the pixel is considered too high to be determined using this method. The extinction maps resulting from this treatment are shown in Fig. \ref{fig_nice-extinction}.


The extinction map of DC303 reveals a high extinction gradient towards the position of the IRAS source, forming a symmetric core projected on the plane of the sky. There is a region of about 20$''$ in radius in which no H band stars are detected. Since at least one star within the FWHM range is required, this results in a few pixels in which the extinction is too high to be determined. These pixels are marked white on the extinction map. The maximum extinction in the map, $A_\mathrm{V}=28^\mathrm{m}$, is located at the pixel closest to the IRAS source.


The extinction in TPN is significantly lower than in DC303 (Fig. \ref{fig_nice-extinction}, right panel). The maximum, $A_\mathrm{V}=10.2^\mathrm{m}$, is located at the coordinates $(ra, dec)$=(12$^h$h44$^m$55\fs6, -78$^d$48$^m$16$^s$). The core appears to be slightly elongated in northwest-southeast direction and the gradient of extinction on the northwest side of the maximum is clearly higher than on the southeast side. The size of the core is quite similar with DC303, both having $A_\mathrm{V} > 3^m$ extending approximately 2$'$ in radius.

   \begin{figure*}
   \centering
   \includegraphics[width=0.95\columnwidth]{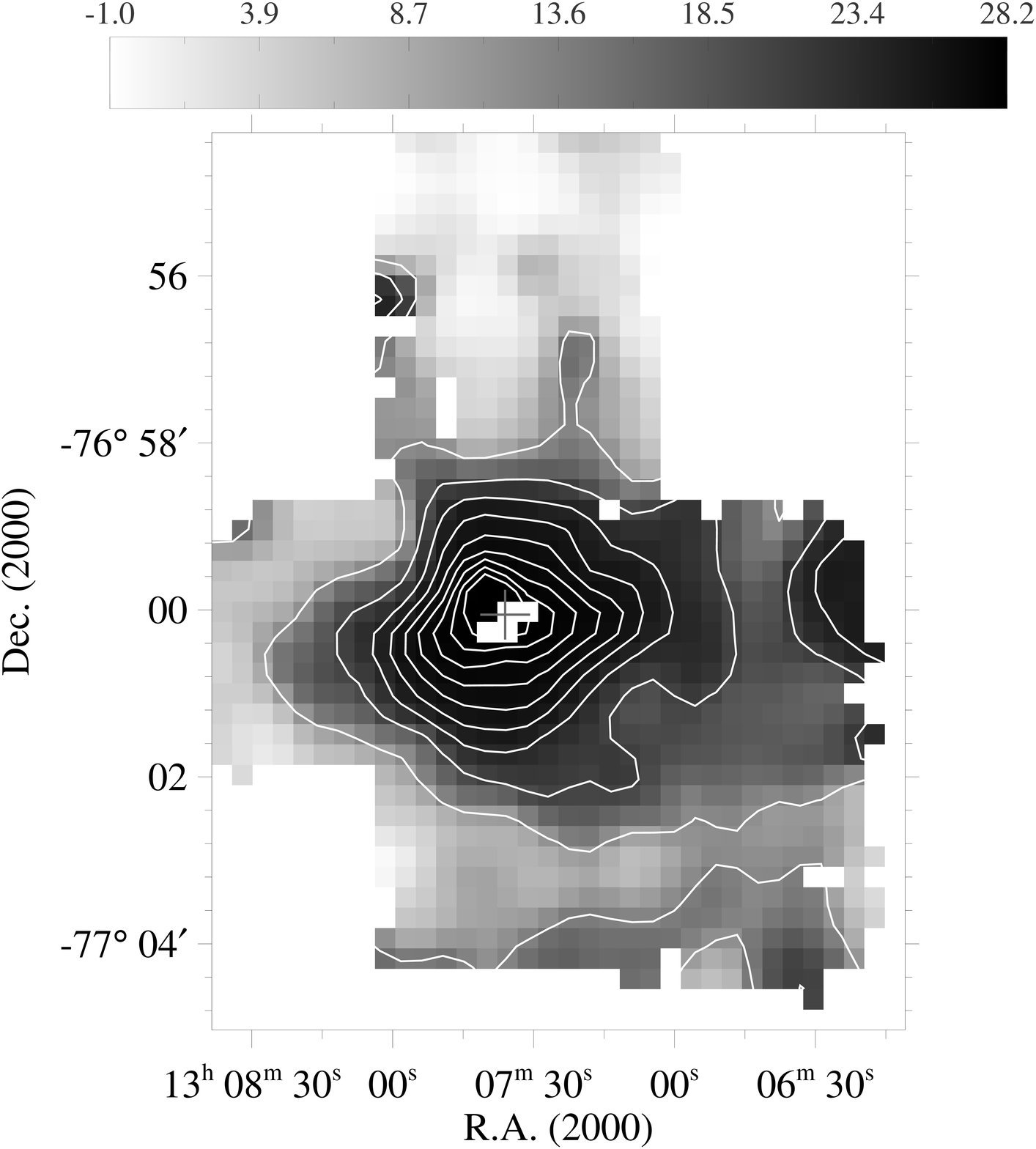} \includegraphics[width=0.95\columnwidth]{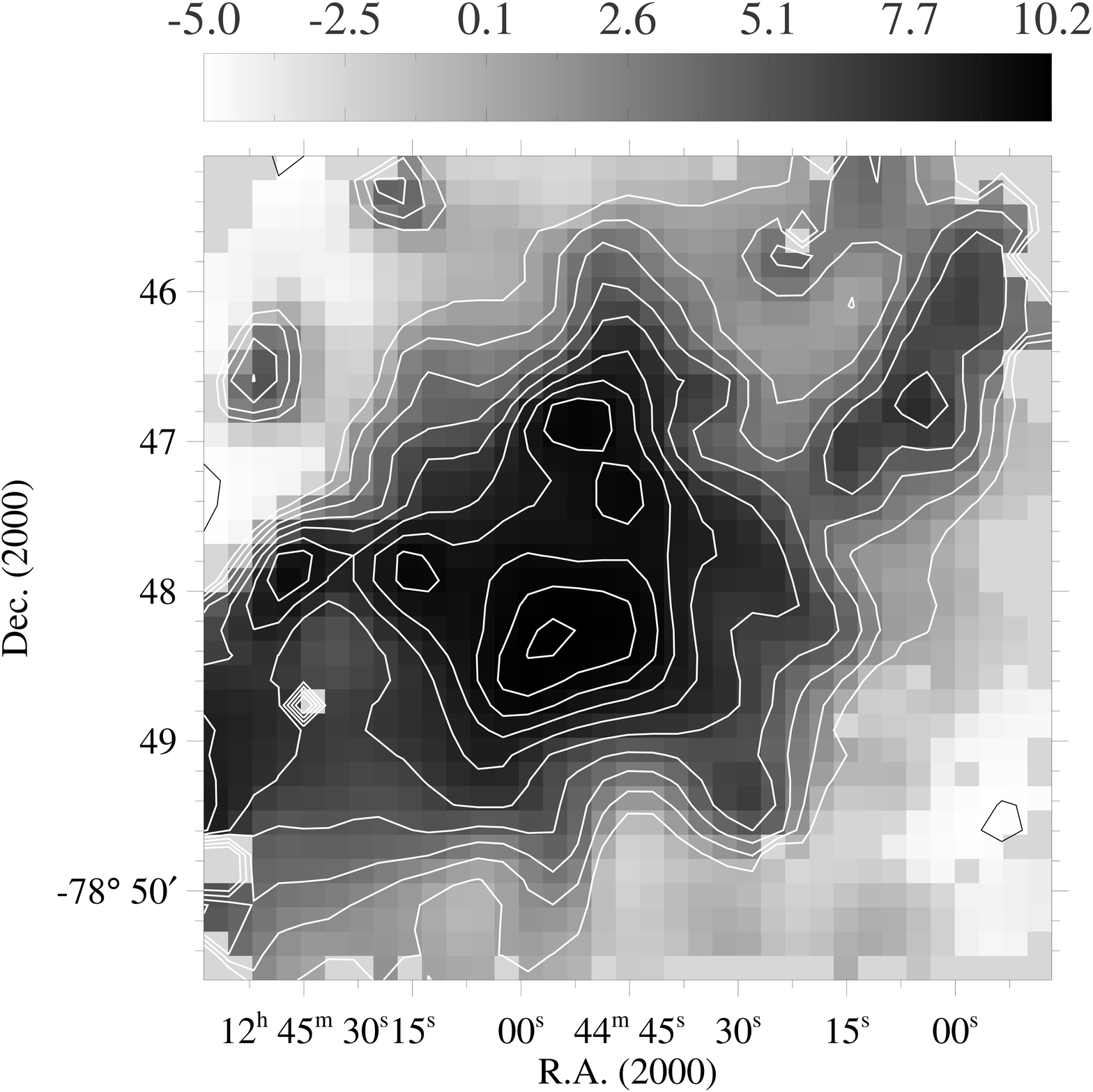}
    \caption{{\bf Left: }The visual extinction map of DC303 derived from the VLT/ISAAC observations using the NICE color excess method. The FWHM resolution of the map is 40$''$. The white pixels result from not having a single star within the FWHM range from the pixel center. The plus sign marks the position of IRAS 13036-7644. Contours start from 3.5 mags and the the step is 2.5 mags.{\bf Right: }The same for TPN. The FWHM resolution of the map is 30$''$. Contours start from 1 mag and the the step is 1 mag.}
         \label{fig_nice-extinction}
   \end{figure*}

   \begin{figure*}
   \centering
   \includegraphics[width=0.95\columnwidth]{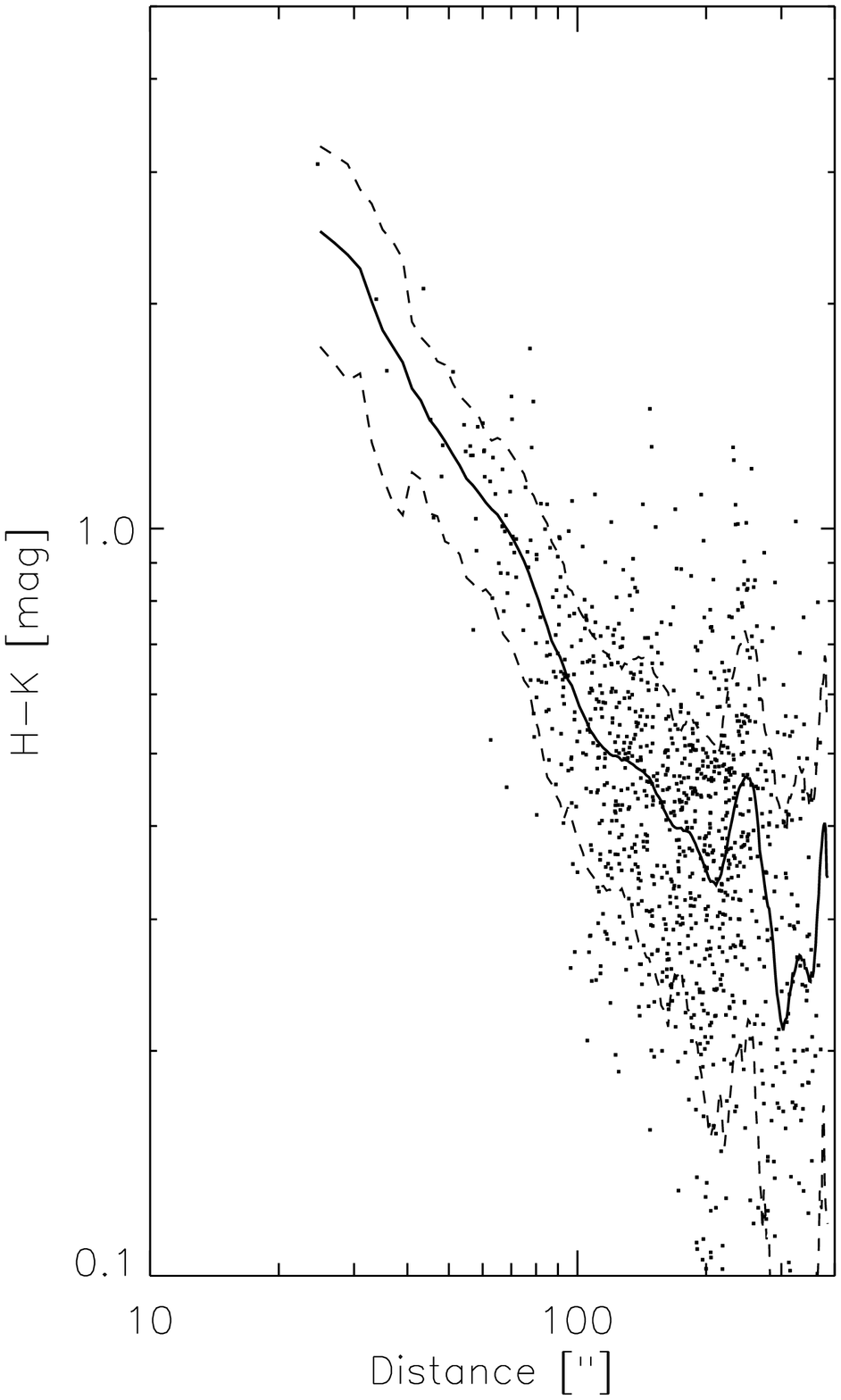} \includegraphics[width=0.95\columnwidth]{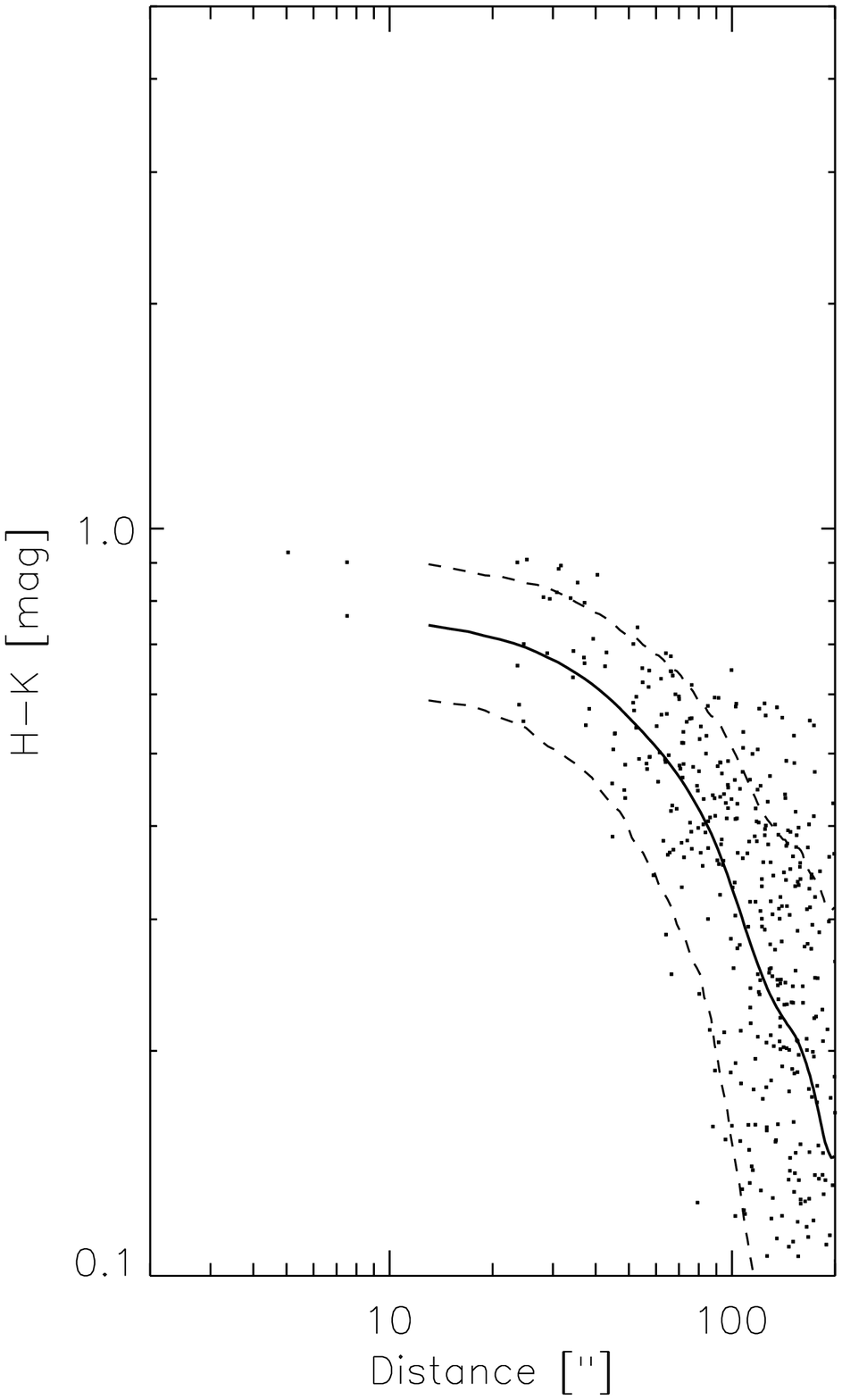}
      \caption{{\bf Left: }The $H-K$ color as a function of the distance from the IRAS point source. The solid line is gaussian convolved average of the colors and dashed lines the standard deviation. {\bf Right: }The same for TPN.}
         \label{fig_color-profiles}
   \end{figure*}


\subsection{Models of radial density distribution}
\label{sec_rho-models}

Fig. \ref{fig_color-profiles} shows the $(H-K_{\mathrm S})$ colors of all the observed stars as a function of distance from the center of the cloud. In DC303 the position of the IRAS source has been selected to represent the center point of the core. In TPN the center point is chosen to be the point of maximum extinction of the $A_\mathrm{V}$ map.
The solid lines in the figures are the gaussian convolved averages of the colors in annuli 20$''$ and 15$''$ wide, for DC303 and TPN respectively. The dashed lines give the standard deviation of the colors. In DC303, the star closest to the IRAS source lies at the angular distance of $r\sim 25$$''$  and possesses the highest ($H-K_{\mathrm S}$) color in the sample, i.e. $(H-K_{\mathrm S})=3.08$ corresponding to about $A_{\mathrm V}\approx 47^m$. In TPN the star closest to the center point is located at the distance of 5$''$ and has the color $(H-K_{\mathrm S})=0.93$ corresponding to the visual extinction of $A_\mathrm{V} \approx 13^m$. 

   \begin{figure*}
   \centering
   \includegraphics[width=0.95\columnwidth]{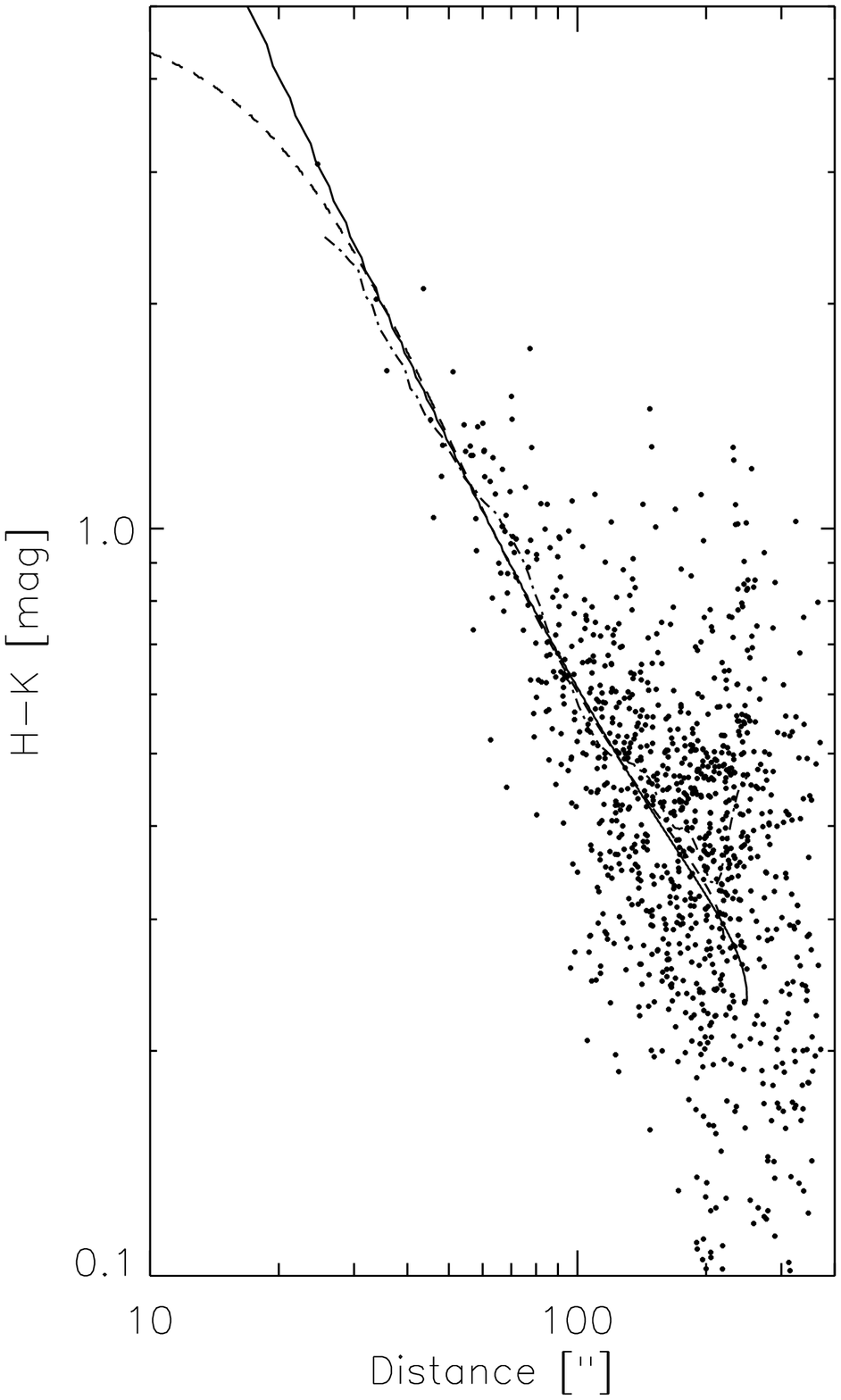} \includegraphics[width=0.95\columnwidth]{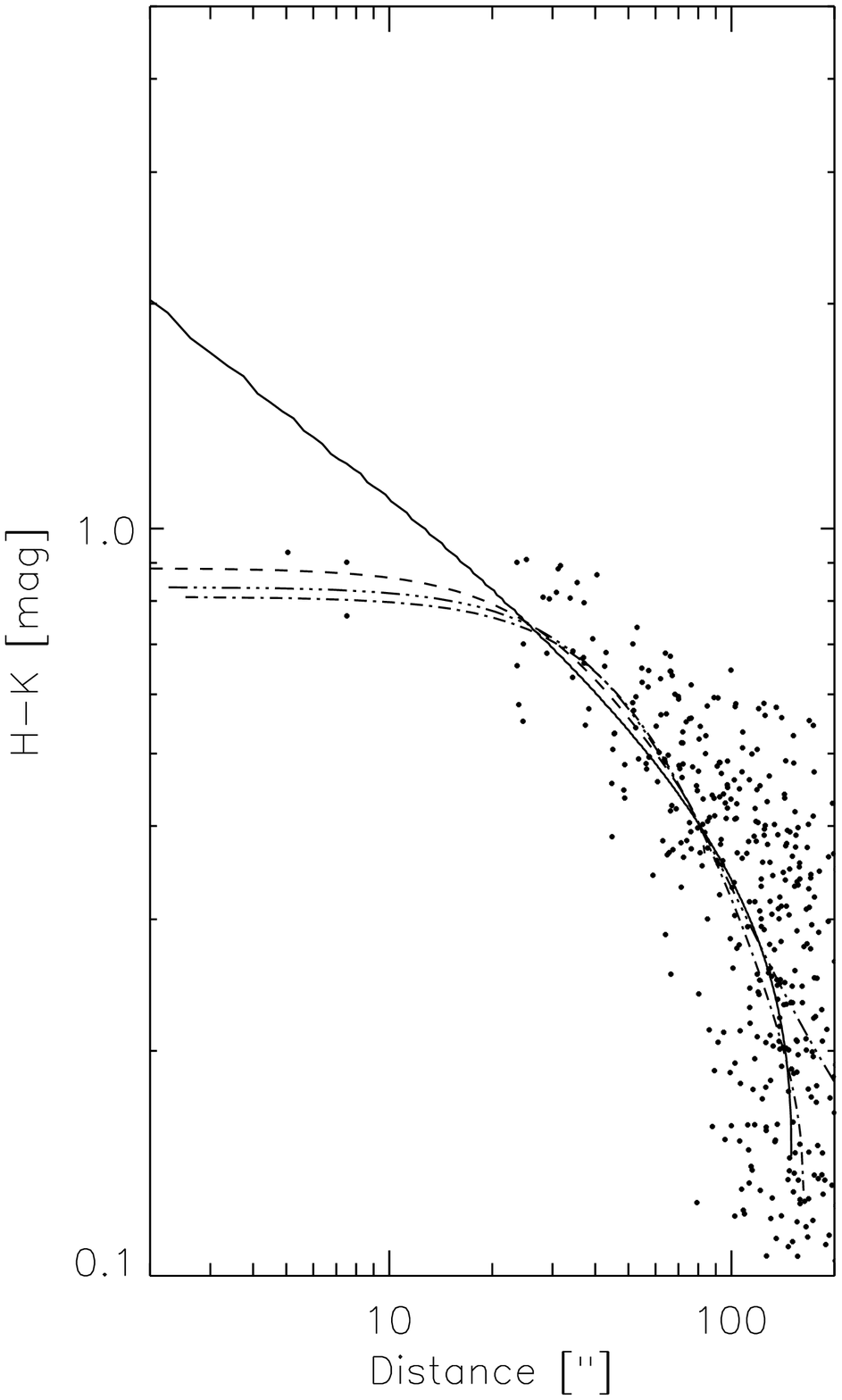}
      \caption{{\bf Left: }The best fitting models of radial density distribution for DC303. The solid line is for a power-law with an exponent $p=2.29\pm 0.08$, the dashed line is a Bonnor-Ebert model with $\xi_{max}=23 \pm 3$. The dash-dotted line is gaussian convolved average. {\bf Right: }The same for TPN. Solid line is the power-law with an exponent $p=1.19\pm 0.09$ and dashed lines Bonnor-Ebert models with different $\xi_{max}$ parameter (see text).}
         \label{fig_color-profiles2}
   \end{figure*}

In the following we model the observed column density profile using two models for radial density distribution: the spherically symmetric inside-out collapse scenario of S77, which essentially means the power-law density distribution, and the Bonnor-Ebert model (Ebert \cite{ebert55}; Bonnor \cite{bonnor56}). The radial density profiles provided by the models are converted to column density profiles and compared to the normalized ($H-K_\mathrm{S}$) data values (Fig. \ref{fig_color-profiles}). To evaluate the goodness of fit the chi-square value between the model profile and the data is used:
\begin{equation}
\label{eq_chi-square}
\chi^2=\frac{1}{dof} \sum \big[ \frac{E(H-K_\mathrm{S})_i^{obs} - E(H-K_\mathrm{S})_i^{model}} {\sigma_i^{obs}} \big] ^2.
\end{equation}
The $dof$ stands for the degrees of freedom in the fit, and $\sigma_i^{obs}$ is the uncertainty related to each observation. The uncertainty is evaluated as a quadratic sum of the photometric uncertainty of the observation, typically 0.08 mag, and the uncertainty in the intrinsic color:
\begin{equation}
\label{eq_sigma}
\sigma_i^{obs}=\sqrt{ \sigma_i^{phot} + \sigma_{bg} }.
\end{equation}

\subsubsection{Inside-out collapse}


The basic scenario of spherical symmetric collapse of cloud cores, the ``inside-out'' collapse, is described by S77. The inside-out collapse begins from the initial stage where the core is relaxed to a balance between gravitation and pressure, and the density distribution of a singular isothermal sphere, $\rho(r) \sim r^{-2}$, is established. The collapse begins in the center of the cloud, and propagates then outwards as a shock front with a speed equivalent to the speed of sound. While the density profile in the outer envelope remains at $\sim r^{-2}$, the density profile of the collapsing region approaches $\rho \sim r^{-3/2}$, indicative of free-fall motion.

The normalization of the density distribution at the static initial condition is determined by the effective sound of speed, $a=(kT/\mu m_H)^{1/2}$, where $T$ is the kinetic temperature and $\mu$ the mean molecular weight. The complete equation for the density distribution has the form:
\begin{equation}
\rho_{static}(r)= \frac{a^2}{2\pi Gr^2}.
\label{eq_sis-normalization}
\end{equation}



The core of DC303 appears to be very symmetric to the extent of about 180$''$. The globule itself is extended mainly westwards from this densest core. Only the stars closer than 180$''$ are included in the fit to avoid the contamination from this more extendend part. 

If the observed extinction profile of DC303 is assumed to originate merely from the core itself, the best-fit power-law has the exponent $p=2.03 \pm 0.04$ $(\chi^2=1.58)$. 
If an additional ``forescreen'' component is included as a free parameter to mimic a more diffuse layer surrounding the core, the fit quality improves resulting in a slightly higher exponent $p= 2.29\pm 0.08$. In the best fit the forescreen component has the value $E(H-K)=0.09$ ($\chi^2=1.54$).
Considering the extinction extending westwards from the core the inclusion of the additional component seems well justified. The power-law fit with a forescreen component is illustrated in Fig. \ref{fig_color-profiles2}. The fit with a forescreen extinction is not significantly dependent of the selected radius of the core. The profile in Fig. \ref{fig_color-profiles2} is calculated with R$_{out}=250''$.

We find that the scaling set by the static condition of S77 (Eq. \ref{eq_sis-normalization}) matches the observed scale of the extinction well. We have used the temperature $T=11$ K evaluated from the C$^{18}$O data (Kainulainen \cite{kainulainen04}), the reddening law of Rieke \& Lebofsky (\cite{rieke85}) and standard gas-to-dust ratio of Bohlin et al. (\cite{bohlin78}, see Eq. \ref{eq_bohlin}).


The extinction profile of TPN at $r>10''$ is well fitted with quite a flat power-law having exponent $p=1.19 \pm 0.09$ $(\chi^2=1.13$). 
However, the three datapoints with $r<10''$ show clearly a turn-over in the profile, and thus a power-law-like model is not really reasonable for TPN. 

\subsubsection{Bonnor-Ebert models}

Recently, pressure-confined, isothermal gas spheres known as ``Bonnor-Ebert spheres'' (BE) have been succesfully used to reproduce the observed column density profiles of dark cloud cores (e.g. Alves et al. \cite{alves01}; Harvey et al. \cite{harvey01}; Harvey et al. \cite{harvey03}; Kandori et al. \cite{kandori05}). The BE spheres are described in detail by Ebert (\cite{ebert55}) and Bonnor (\cite{bonnor56}). As the singular isothermal model of S77 the BE spheres satisfy the Lane-Emden equation:
\begin{equation}
\label{eq_lane-emden}
\frac{1}{\xi^2} \frac{d}{d\xi} \big( \xi^2 \frac{d\phi}{d\xi} \big) = \mathrm{exp}[-\phi],
\end{equation}
where $\xi$ is the dimensionless radius and $\phi$ is the logarithmic density contrast: 
\begin{equation}
\phi(\xi)=-\ln{\frac{\rho}{\rho_c}}.
\end{equation}
The dimensionless radius is linked to the physical radius through equation:
\begin{equation}
\label{eq_dimensionless-radius}
\xi = \frac{r}{R_0}= \big( \frac{r}{a} \big) (4\pi G\rho_c)^{\frac{1}{2}},
\end{equation}
where $R_0$ is the ``scale radius'' and $\rho_c$ the central density. The shape of each BE sphere, i.e. the solution of Eq. \ref{eq_lane-emden}, can be characterized by a single parameter, the outer radius of the sphere in dimensionless units: $\xi_{max}$. The comparison of the observed profile with the one derived by solving Eq. \ref{eq_lane-emden} thus fixes the parameter $\xi_{max}$ and the shape of the density distribution. 

Bonnor (\cite{bonnor56}) examines the stability of the solutions and concludes that solutions with $\xi_\mathrm{max}$ value larger than 6.5 are unstable and susceptible for collapse. The solution with $\xi_\mathrm{max}=6.5$ represents the marginally stable configuration. The $\xi_\mathrm{max}$ value is directly linked to the density contrast: for example, the value $\xi_\mathrm{max}=6.5$ corresponds to the center-to-edge density contrast $\sim$14 and the value $\xi_\mathrm{max}=10.0$ to $\sim$42.

While the shape of the density profile is fixed by determining the parameter $\xi_\mathrm{max}$, matching the \emph{scale} of the fitted profile to the observed profile requires fixing of two additional parameters in Eq. \ref{eq_dimensionless-radius}. The common way to do this is to fix the soundspeed using temperature information derived from e.g. molecular line observations, and physical radius of the fitted sphere by assuming a certain distance to the globule. However, as discussed in Harvey et al. (\cite{harvey03}) assuming additional constraints such as the outer edge can result in underestimation of the systematic errors. A more robust fit is achieved by fitting the scale radius, $R_0$, instead of the dimensionless radius, $\xi_\mathrm{max}$ (see Eq. \ref{eq_dimensionless-radius}). In the following we follow this procedure and fit BE models for DC303 and TPN.


The BE model that best fits the shape of the observed column density profile of DC303 is described by the parameter $\xi = 23 \pm 3$. As in the case of power-law fitting the inclusion of a constant term improves the fit quality significantly. The model with a constant $(H-K)_\mathrm{fg}=0.14$ extinction forescreening the BE sphere reduces the $\chi^2$ of the fit from 1.60 to 1.47.

Fitting the scale radius of the BE model did not provide a robust fit for TPN, most likely due to non-circular projection and the small coverage of the profile compared to the uncertainties in observed colors. The confusion from the outerparts of the globule becomes significant already at $\sim 80''$ distance which makes the profile quite noisy. The best fit is marginally achieved with a profile having $\xi_{max}= 13 \pm 3$, but essentially any profile having $\xi_{max}>8$ produces equally good fit with $\chi^2$ changing less than one percent. The Fig. \ref{fig_color-profiles} shows the fitted profiles with $\xi_{max}=$8, 12 and 16. Interestingly, the inclusion of an additional extinction layer results in a robust fit and, in fact, produces similar solution for a density distribution with $\xi_{max}= 10 \pm 2$. However, including the forescreen extinction does not seem reasonable in this case, as there is no component of extinction extending beyond the radius of the BE model. Thus we conclude that we can only give lower limit for the $\xi_{max}$ parameter in the case of TPN. 

\begin{table*}
\caption{The summary of fitted models}             
\label{tab_model-fitting}      
\centering          
\begin{tabular}{l c c l }  
\hline\hline       
Model     & Region &  $\chi^2$   & Parameters \\ 
\hline
DCld303.8-14.2 & & & \\
\hline                    
Single power-law   & $24''<r<180''$   & 1.58 & $p=2.03 \pm 0.04$ \\
Single power-law with forescreen & $24''<r<180''$ & 1.54 & $p = 2.29\pm 0.08, E(H-K)_\mathrm{0} = 0.09$ \\
BE & $24''<r<220''$ & 1.47 & $\xi_{max}=23 \pm 3 $, $E(H-K)_0=0.14$  \\ 
\hline
Thumbprint Nebula & & & \\
\hline
Single power-law   & $23''<r<100'' $& 1.13 & $p=1.19 \pm 0.09$  \\
BE & $6''<r<120''$ & 1.11 & $\xi_{max} > 8 \pm 2 $ \\
\hline                  
\end{tabular}
\end{table*}

\subsubsection{Masses of the globules}

We calculate the masses of the two globules from the extinction maps by summing all the pixels within a certain radius together. We adopt the distance of 150 pc derived by Knude \& H\o g (\cite{knude98}) from the data of the Hipparcos satellite. The gas-to-dust relation derived by Bohlin et al. (\cite{bohlin78}) gives the ratio of H$_2$ column density and extinction:
\begin{equation}
\frac{N(\mathrm{H}_2)}{A_\mathrm{V}} = 9.4 \times 10^{20} \mathrm{\, cm}^{-2}\mathrm{mag}^{-1}.
\label{eq_bohlin}
\end{equation}
Using the equation above, and adding 30 \% due to helium, the masses of the cores are:
\begin{equation}
M = 5.9 \pm 1.8 \mathrm{\,M}_\odot \qquad r<3.4' \mathrm{\,\,(0.15\,pc)}\qquad \mathrm{(DC303)} 
\end{equation}
\begin{equation}
M = 3.7 \pm 1.2 \mathrm{\,M}_\odot  \qquad r<3.4' \mathrm{\,\,(0.15\,pc)}\qquad \mathrm{(TPN)} 
\end{equation}
The masses derived for the cores from the extinction measurement are in good agreement with the masses derived from the C$^{18}$O observations (Kainulainen \cite{kainulainen04} for DC303; Lehtinen et al. \cite{lehtinen95} for TPN).


\section{Discussion}
\label{sec_discussion}

 
The studies of radial density distributions of dark globules exploiting dust emission have established the framework that pre-protostellar cores have shallower density profiles and are less centrally condensed than protostellar cores. Moreover, the column density profiles of the majority of the cores, with or without central object, seem to be well reproduced with density distributions of isothermal Bonnor-Ebert spheres. Despite these findings which are made by comparing samples of tens of globules, the dust emission is not an ideal tool for characterizing the density distributions of individual cores due to interpretational problems and limited beam size.

The extinction measurement using the color excesses of background stars is an effort to overcome these uncertainties and it provides a direct tracer of mass distribution via high signal-to-noise extinction maps. 
 The few previous studies utilizing the technique in high resolution, this paper included, have strengthened the general results of the sub-mm studies: the pre-protostellar cores possess density distributions with flattened inner regions.

However, somewhat contradicting the results from sub-mm observations, the fitted BE models have often indicated unstable configurations even for pre-protostellar cores (e.g. Kandori et al. \cite{kandori05}). If the starless globules are regarded as stable, hydrostatic configurations, there must be a significant additional forces supporting the cloud. Alternatively, the wide range of observed BE solutions can be regarded as an evolutionary sequence of slowly collapsing sphere, in which the object is initially close to equilibrium and slowly becomes more centrally condensed before the actual formation of the central object. In this case, the observed central concentration, or the $\xi_{max}$ parameter, would be directly linked to time since the beginning of contraction (Aikawa et al. \cite{aikawa05}; see also Fig. 8 of Kandori et al. \cite{kandori05}). 

It is necessary to mention that even though the observed \emph{column density} distributions are reproduced using the \emph{density distribution} of the Bonnor-Ebert model, the reality is evidently more complex. Even though the extinction features are well reproduced, the interpretation of the physical conditions or evolution of the core is not straightforward nor uniquely determinable from the extinction alone. The interpretation is hampered by the unknown contributions from additional supporting mechanisms such as magnetic fields and turbulent motions and by the basic assumption of thermodynamic equilibrium which is not likely met, as well as by deviations from spherical symmetry and non-isothermality. These issues are addressed in detailed discussion in Hotzel et al. (\cite{hotzel02}) and Ballesteros-Paredes et al. (\cite{ballesteros03}). In their work Ballesteros-Paredes et al. argued that large fraction of cores which form through turbulent motions might resemble the appearance of BE spheres even though they are not static but rather dynamically evolving objects. 
It should be emphasized that even though the dust extinction technique can trace the column density distribution in high detail, the combination of dust emission, extinction and molecular line observations is required to make the final arguments about the physical properties of the cores.

\subsection{The star forming globule DC303}

In comparison to pre-protostellar globules, which are presumably close to hydrodynamic equilibrium, it is useful to consider globules which are evolved to a stage in which the central object is established and the system is rapidly accreting mass from the envelope (Class 0 YSOs). In the context of dust extinction studies, only few occasions which consider cores in similar stage can be found. Those are: \object{B335} (Harvey et al. \cite{harvey01}), \object{CB188} (Kandori et al. \cite{kandori05}) and \object{L1014} (Huard et al. \cite{huard06}). The column density distributions of these three cores can be fitted with highly unstable BE solutions having dimensionless radii $\xi_{max}>15$. Particularly, L1014 appears to have a very high degree of central concentration with $\xi_{max}$ in the range $29.2>\xi_{max}>36.9$.


The power-law density distribution with $p\sim 2$ derived for DC303 in this study is similar to that of the globule B335, apart from the fact that in B335 the evidence of flattening inner region at $r\lesssim0.03$ pc was found. Harvey et al. (\cite{harvey03}) concluded that the observed extinction features of B335 are well reproduced with a broken power-law density structure with infall radius of S77 model at 26$''\pm 3''$ (0.03 pc at 250 pc), and a simple cone-like outflow model. The position of infall radius matches well the molecular line observations and the results of radiative transfer modeling following the inside-out collapse framework (Zhou et al. \cite{zhou93}; Choi et al. \cite{choi95}). In these studies the age estimate of $\sim 1.3 \times 10^5$ is derived for the central object.

If we perform a simple calculation assuming that the shock-front of the inside-out collapse is moving outwards from the center of DC303 with the speed of $a\sim 200$ m s$^{-1}$, we can place an upper limit of $\sim 1.8 \times 10^5$ for the age of the central object. Here we assume that we should be able to disentangle the flattening from the observations, should it occur at radius higher than $\sim 50''$ (0.036 pc).

B335 is one of the best-known objects harboring a Class 0 protostar (see e.g. Hodapp et al. \cite{hodapp98}). Based on the spectral energy distribution of the young stellar object in DC303, it is presumably in an evolutionary stage where much of the mass of the envelope has already accreted to the protostar/-disk system. Lehtinen et al. (\cite{lehtinen05}) describes the core to be ``a transition object between Classes 0 and I''. This further indicates that DC303 is in a similar stage with B335. The main properties of these globules are briefly quoted in table \ref{tab_DC303-and-B335}. The largest difference between the properties seems to be the estimated total mass, which is more than double for B335 than for DC303. 


It is interesting that even though DC303 clearly does not represent an initial stage of star formation, the normalization of the density distribution is well described by the static condition of the inside-out collapse. This is in contradiction with studies of B335 and L694-2 (Harvey et al. \cite{harvey01}, \cite{harvey03}) where non-explainable scaling factors of $\sim 3-5$ were needed to match the scale of the density distribution to the observed color excess.


The column density profile of DC303 can be equally well fitted with a supercritical BE model. The similarity of these models originates from the fact that the density of a BE model at large radius closely resembles the $r^{-2}$ distribution. At smaller radii the profile flattens. As the observed ($H-K$) colors do not show any flattening the BE model produces effectively a power-law distribution. Even though the BE model is quite succesfull in describing the density distribution, it should be stressed that DC303 hardly resembles the configuration of the BE sphere: the core has already undergone a collapse forming a central object.

\begin{table*}
\begin{minipage}[t]{\columnwidth}
\caption{The properties of DC303 and B335}             
\label{tab_DC303-and-B335}      
\centering  
\renewcommand{\footnoterule}{}  
\begin{tabular}{l c c c }  
\hline\hline       
           & DC303     & B335      & Comments \\ 
\hline
YSO        & Class 0/I & Class 0   &  1, 2   \\
Age        & $<1.8\times 10^5$ a & $1.3\times 10^5$ a & 3, 4 \\
$M$        &  5.9 M$\odot$ & 14 M$\odot$ & 3, from NIR extinction \\
$M_\mathrm{central \, object}$ & - & 0.37  M$\odot$  & 4 \\
$T_\mathrm{kin}$ & 11 K & 13 K & 5, 4 \\

\hline                  
\end{tabular}
\flushleft
References: (1) Lehtinen et al. \cite{lehtinen05} (2) Hodapp et al. \cite{hodapp98} (3) This study (4) Zhou et al. \cite{zhou93} (5) Kainulainen \cite{kainulainen04}

\end{minipage}
\end{table*}


\subsection{The starless globule TPN}

Unlike DC303, TPN harbors no embedded source and is in a very different stage of evolution. The molecular line observations have not shown features indicative for collapse or outflow, but have proven the core to be very cold, $T_{ex}\approx 6$ K, and close to hydrostatic equilibrium (Lehtinen et al. \cite{lehtinen95}, \cite{lehtinen05}). No significant nonthermal turbulence or large scale motions were found in the studies. It should be justified to conclude that collapse has not taken place, nor is currently taking place in the core. The maximum extinction derived for TPN, $A_\mathrm{V}\approx 13^m$, is also relatively low taking into account that the distribution is resolved to a very small radial distance (6$''$, corresponding to $\sim 1000$ AU at 150 pc). Correspondingly, the mass of the TPN, $3.7 M_\odot$, is quite low compared to globules for which the mass has been determined using the same method (e.g. Kandori et al. \cite{kandori05}, Table 5).


The BE-model of TPN was not conclusively determined, but we note the lower limit for the parameter $\xi_{max}\gtrsim 8$. The projection of TPN also deviates somewhat from spherical symmetry, which makes the application of BE models more questionable. The derived lower limit implies an unstable configuration disagreeing with the characteristics derived from molecular line observations. However, this is consistent with the tendency that a large fraction of globules that are assumed to be starless are nevertheless best fitted with unstable BE-models (e.g. Kandori et al. \cite{kandori05}). In the context of slowly contracting BE-sphere (Aikawa et al. \cite{aikawa05}) TPN would be in very early stage of contraction with $t\sim 5\times 10^5$ years since the stable configuration was actual. There is no detections of infall motions in TPN, but the infall speeds at early stages would be too low to be detected from the current data. Alternatively, the high central concentration suggests the presence of supporting forces in addition to thermal support, and the possibility that TPN is not contracting but rather close to equilibrium. With the current data it is not possible to conclude whether TPN is slowly contracting towards the star formation, or close to equilibrium and perhaps even dissolving in the future. 


\section{Conclusions}
\label{sec_conclusions}

We present deep near-infrared observations of globules DC303 and TPN made with the ISAAC instrument on the ESO/VLT telescope. The $H-K$ colors of the observed stars are used as a measure of dust column density and the observed column density profile is examined in the light of theoretical models. The main conclusions of our work are as follows:

\begin{enumerate}

\item The $J$, $H$ and $K_\mathrm{S}$ band frames of DC303 observed with ISAAC show a dense core with peak extinction higher than $A_\mathrm{V}>50^m$ surrounding the embedded IRAS source. The data of TPN show a more transparent core with peak extinction of about $A_\mathrm{V}\sim 13^m$. From DC303 the images also reveal Herbig-Haro-like objects close to the core center, and a ring-like ``halo'' structure of scattered light surrounding the core.

\item The best-fit power-law describing the density distribution of DC303 has an exponent $p=2.29\pm 0.08$ and an additional $H-K=0.09$ layer of extinction. If the additional extinction layer is not included, the power-law exponent has the value of $p=2.03 \pm 0.04$. The density distribution shows no signs of flattening towards the core center ($r>0.03$ pc). The normalization of the power-law models is well fitted with the static condition of singular isothermal model of S77 and the normal reddening law. The density distribution of TPN cannot be well fitted with a single power-law.

\item For DC303, the Bonnor-Ebert model with a dimensionless radius $\xi_\mathrm{max}=23 \pm 3$ offers an equally good fit to the shape of the observed column density profile as the power-law model. The radial density distribution of TPN differs greatly from that of DC303 as it shows flattening of the profile at the distance of about 0.015 pc from the core center. We are unable to describe TPN consistently with a Bonnor-Ebert model, but derive the lower limit for the dimensionless outer edge: $\xi_\mathrm{max}\gtrsim 8$.

\end{enumerate}


\begin{acknowledgements}

J. K. acknowledges the support of the Academy of Finland, grant No. 206049. L.B. acknowledges support from the Chilean Centro de Astrof\'isica FONDAP No. 15010003.

\end{acknowledgements}

\end{document}